\documentclass{aa}
\usepackage{graphicx}
\usepackage{natbib}
\bibpunct{(}{)}{;}{a}{}{,} 

\begin{document}

\newcommand{\cen}{\centerline}
\newcommand{\noin}{\noindent}
\newcommand{\kms}{km\,s$^{-1}$}

\newcommand{\gala} {\object{NGC 3184}\,}
\newcommand{\galb} {\object{NGC 3938}\,}

\title{Density effect on multiwavelength luminosities on star-formation regions  in \gala and \galb}
\author{{ A. Cald\'u-Primo\inst{1}}
\and {I. Cruz-Gonz\'alez\inst{1}}
\and {C. Morisset\inst{1}}}

\offprints{I. Cruz-Gonz\'alez, \email{irene@astroscu.unam.mx}}

\institute{Instituto de Astronom\'{\i}a, Universidad Nacional Aut\'onoma de M\'exico, M\'exico D.~F., M\'exico}

\date{Received 25 June 2008 / Accepted 02 Oct. 2008}

\abstract {}  {We analyzed the regions of star formation in the
spiral galaxies \gala  and \galb from archive images for a wide
range of wavelengths (NUV from GALEX, H$\alpha$ from JKT and KPNO,
8 and 24 $\mu$m from Spitzer, and CO from BIMA).} {We used the
Clump Find Algorithm to extract the properties of the star-forming
tracers, identifiable as emission regions at each wavelength.} {We
obtained a power-law relation between the luminosity and the
emission region volume that scales as expected, $L \propto V$, for
the H$\alpha$ and NUV emission, but the luminosity varies far more
rapidly with the volume for the dust (8 and 24 $\mu$m) and
molecular gas emitting regions in CO. This is interpreted as a
change in the emissivity  with the size of the cloud, either by an
augmentation of the overall density or due to the presence of high
density clumps, with high local emissivity coefficients. Although
the clumpy nature of molecular gas may be unsurprising, the clumpy
nature of mid-infrared emission regions, which could be explained
by newly formed high to intermediate mass stars embedded in the
dust providing the heating, is clearly revealed in both
galaxies.}{}


\titlerunning{Density effect on SF luminosities in \gala and \galb}

\authorrunning{A. Cald\'u-Primo et al.}

\maketitle

\section{Introduction}

Surveys completed with both the Spitzer Space Telescope
\citep{werner04} and the Galaxy Evolution Explorer
\citep[GALEX;][]{martin05} provided morphologically detailed IR
and UV images of galaxies. Likewise, surveys such as the Berkeley
Illinois Maryland Association (BIMA) Survey of Nearby Galaxies
\citep[SONG;][]{helfer03} provided us with aperture synthesis CO
mapping surveys of galaxies. These surveys are enabling
multiwavelength studies of star-formation activity with adequate
resolution and sensitivity
\citep[e.g.][]{calzetti05,calzetti07,kennicutt07,thilker07}, and
the cross calibration of star-formation rate (SFR) indicators at
different wavelengths \citep[see][]{calzetti07}.

Since the work by \citet{kennicutt88}, studies of  \ion{H}{II}
regions in spiral galaxies by means of their H$\alpha$ emission
have demonstrated that the logarithm of the diameter of the
\ion{H}{II} region and the logarithm of the H$\alpha$ luminosity
scale roughly with a value of one third, i.e. linearly with the
volume, which is the slope expected for constant density
radiation-bounded nebulae. This trend is found in studies of
\ion{H}{II} regions of several spiral galaxies, such as
\object{NGC 7331} \citep{marcelin94}, \object{NGC 3992}
\citep{cepa89},  and \object{NGC 4321} \citep{cepa90a}, as well as
in irregular galaxies such as \object{NGC 6822} \citep{hodge89},
and \object{IC 10} \citep{hodge90}.

We present a study of star-formation regions in the spiral face-on
galaxies \gala and \galb using images at various wavelengths (NUV
from GALEX, H$\alpha$ from JKT and KPNO, 8 and 24 $\mu$m from
Spitzer, and CO from BIMA). Our aim is to explore the possible
slope values of the relation $L_{\nu} \propto V^{\beta}$ for the
collection of individual gaseous or dusty star-forming regions of
volume $V$ and luminosity $L_{\nu}$ detected at each frequency
bandwidth $\nu$.

This work is a phenomenological study to provide clues in
understanding the different processes affecting the observed beta
slope.

\section{Multiwavelength Images of \gala and \galb}

The galaxy images were taken from archive data. We used the
Spitzer Infrared Nearby Galaxy Survey
\citep[SINGS;][]{kennicutt03} database to select candidate spiral
galaxies oriented nearly face--on, which are the most suitable for
our study because optical and UV data are less affected by
extinction. From a short list of candidates, we selected \gala and
\galb, both with available 8 $\mu$m Infrared Array Camera
\citep[IRAC;][]{IRAC} images and  24 $\mu$m  Multiband Imaging
Photometer \citep[MIPS;][]{MIPS} images from Spitzer with a
resolution of 0.75\,\arcsec\,pixel$^{-1}$. For both galaxies, CO
maps at 2.6 mm (J:1$\rightarrow$0) are also available with a
resolution of 0.997\,\arcsec\,pixel$^{-1}$ $\times$
1.009\,\arcsec\,pixel$^{-1}$ from BIMA SONG \citep{helfer03}. The
H$\alpha$ image of \gala\ with a resolution of
0.234\,\arcsec\,pixel$^{-1}$ is from the archive of the Jakobus
Kapteyn Telescope of the Isaac Newton telescopes \citep{knapen04},
and that of \galb\ was obtained from the NASA/IPAC Extragalactic
Database (NED) taken at the 2.1m telescope at Kitt Peak National
Observatory (KPNO) of NOAO (National Optical Astronomy
Observatory) with a resolution of
0.303\,\arcsec\,pixel$^{-1}\,\times\,$0.302\,\arcsec\,pixel$^{-1}$.
The 2271 \AA\ (NUV) image was obtained by GALEX \citep{martin05},
and has a resolution of
1.513\,\arcsec\,pixel$^{-1}\,\times\,$1.514\,\arcsec\,pixel$^{-1}$,
and covers a bandwidth of 1750-2800 \AA.

\section{Clump Finding Analysis}

To be able to study the clumpy structure of the galaxies at each
wavelength, we used the Clump Finding algorithm developed by
\citet{williams94} to study structure in molecular clouds.  We
have used a 2D version of the routine, Clumpfind2D,  originally in
Fortran and then adapted to IDL. This algorithm defines different
clumps using the ``friends-to-friends'' routine, which groups
pixels with some sort of connectivity among them, and inside a
specific range of intensity. The final result of the Clumpfind2D
routine is a list of the different clumps, their position, and
their area in pixels.

\begin{figure}[tp2]
  \centering
{\includegraphics*[width=\columnwidth]{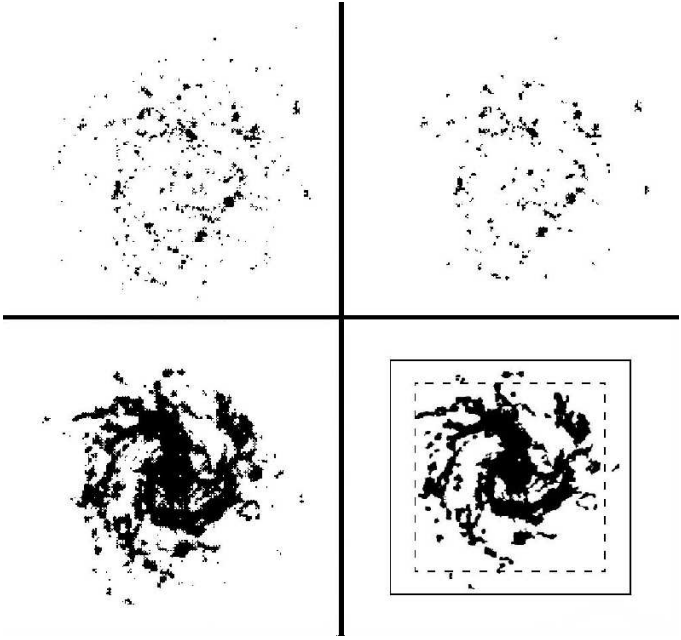}}
  \caption{\gala. Comparison of the original images (left) and those obtained with
  the Clumpfind2D algorithm (right) at H$\alpha$ (top) and 8 $\mu$m (bottom).
  Both images cover an 8\,\arcmin\,$\times$\,8\,\arcmin region, while in the bottom
  right panel the solid line square is the area studied in CO (5.7\,\arcmin\,$\times$\,5.6\,\arcmin)
  and the dashed square is the 24 $\mu$m image (4.5\,\arcmin\,$\times$\,4.5\,\arcmin).}
\label{fig1}
\end{figure}

Galaxy images were preprocessed. Foreground stars were removed
from the original galaxy images  using standard IRAF routines. The
areas explored for clumps in each galaxy correspond to the minimum
area in which the galaxy is observed at all wavelengths. For
\gala, the areas studied are H$\alpha$
8$\,\arcmin\,\times\,$8\,\arcmin, 8 $\mu$m
8$\,\arcmin\,\times\,$8\,\arcmin, 24 $\mu$m
4.5$\,\arcmin\,\times\,$4.5\,\arcmin, and CO
5.7$\,\arcmin\,\times\,$5.6\,\arcmin (c.f. bottom right panel in
Fig.~\ref{fig1}). For \galb, a similar area of
5.7$\,\arcmin\,\times\,$5.6\,\arcmin\ was studied at all
wavelengths, NUV, H$\alpha$, 8 and 24 $\mu$m, and CO.

As an example of how the morphology is well reproduced, in
Fig.~\ref{fig1} we show the original image of  \gala\ at H$\alpha$
(top) and 8 $\mu$m (bottom) with the corresponding images obtained
by Clumpfind2D (right panel).

\section{$L_{\nu} \propto V^{\beta}$}

The flux-calibrated original image and the clump or
emission-region parameters obtained with ClumpFind2D allow us to
calculate the luminosity of each region emitted at a particular
wavelength. To calculate emission-region luminosities, we assumed
a distance of 11.5$\pm$2.0 Mpc for \gala\ and of 11.6$\pm$0.8 Mpc
for \galb, taken from NED. Different corrections were applied to
the calibrated fluxes depending on the wavelength, for NUV, a
Milky Way \citep{cardelli89} and internal extinction
\citep{buat05} corrections; for H$\alpha$, a Milky Way extinction
\citep{cardelli89} and contribution of  \ion{N}{[ii]}
\citep{calzetti07} corrections; and for 8 $\mu$m, a Milky Way
extinction correction \citep{indebetouw05}. No corrections were
made for CO and 24 $\mu$m fluxes. Luminosities were calculated at
the corresponding frequencies of H$\alpha$, 8 and 24 $\mu$m, CO
2.6 mm, and NUV (2271 \AA) if available. We therefore calculate
luminosities of regions of emitting gas (ionized or molecular) or
dust at 8 and 24 $\mu$m, which have been identified as
characteristic tracers of star-forming regions.

We noticed a relation between the areas of the emission regions
and their luminosities for a given wavelength. Assuming a
spherical geometry, the volume was found based on the equivalent
radius obtained from the original emission-region area. The
relations between the multiwavelength luminosities $L_{\nu}$ of
the clumps and their corresponding volumes $V$ (defined to be $4
\pi R^3/3$, where $R$ is the mean radius of the emission region)
obtained for \gala\ and \galb\ are of the form $L_{\nu} \propto
V^{\beta}$.

\begin{figure*}[tp3]
\begin{center}
 {\includegraphics[width=17cm]{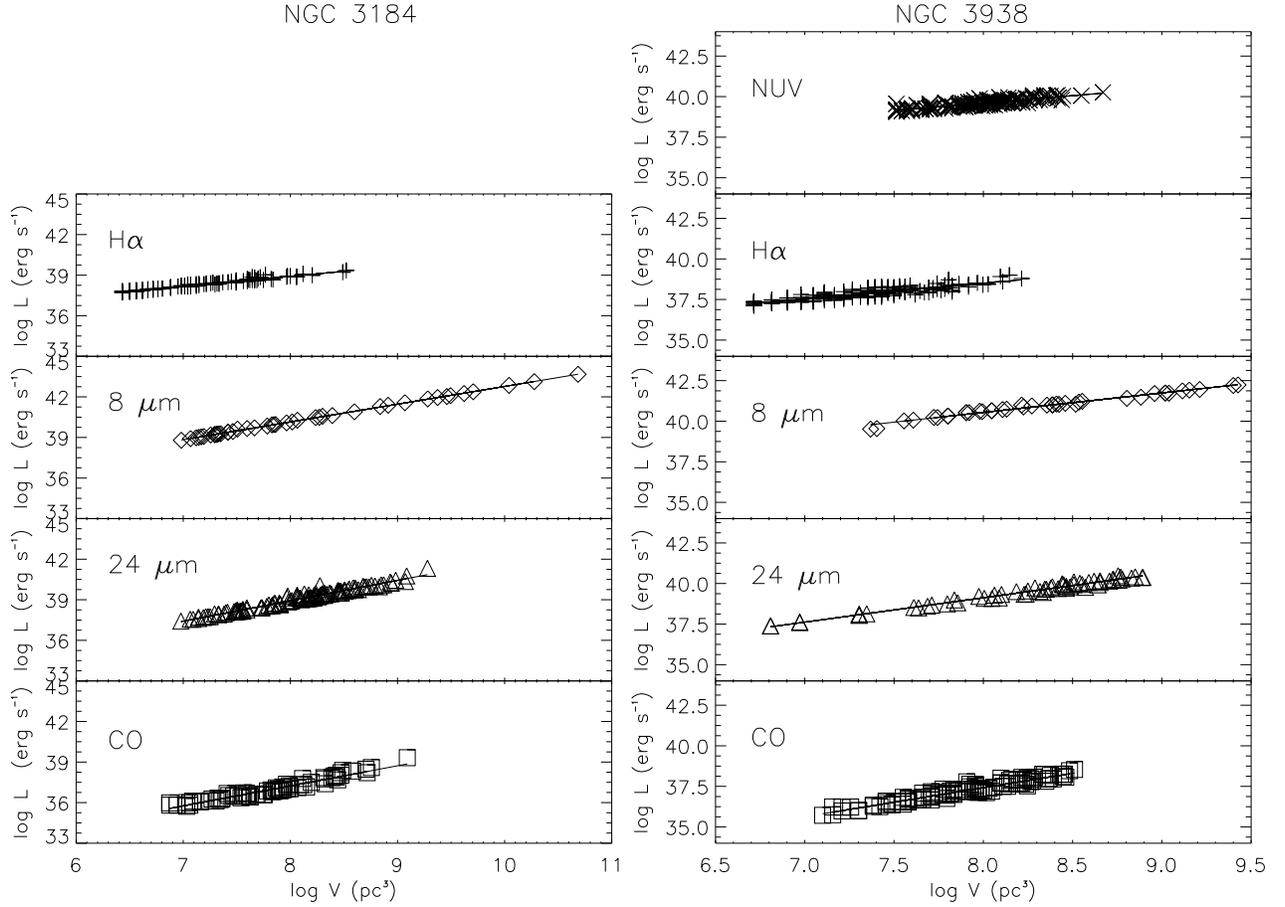}}
 \caption{Relation between the luminosity and volume of the
 emission region obtained for different star formation tracers in \gala and \galb.}
 \label{fig2}
 \end{center}
\end{figure*}

To visualize the difference between all of star-formation tracers,
we present them in different panels on Fig.~\ref{fig2} and plotted
together on Fig.~\ref{fig3}. In both galaxies, the dispersion in
H$\alpha$ is comparable to previous results
\citep[e.g.][]{kennicutt88}. For CO and NUV, the dispersion is
comparable with that in H$\alpha$, while for both 8 and 24 $\mu$m
the dispersion is smaller.

For \galb, the slope values ($\beta$) for both H$\alpha$ and NUV
are close to 1, while for the other tracers the slopes are much
steeper (see the right panel of Fig.~\ref{fig2}). On the other
hand, for \gala the slope for H$\alpha$ does not become close to
1, but the other luminosities follow the same trend with volume as
that found in \galb. For both galaxies, the dust tracers render
the most similar $\beta$ values, being identical for 24 $\mu$m.
The steepest value is obtained for the two galaxies with the CO
clumps.

\section{Discussion}

\subsection{Star-formation tracers}

The various wavelengths are related to the star-formation process
in different ways. The NUV emission originates in regions
populated by massive, young stars (ages $\leq 10^{8}$ yr). This
radiation is affected significantly by extinction, but after the
necessary corrections have been made, its intensity relates
directly to the actual SFR \citep{iglesias06}. The H$\alpha$
emission is the principal recombination line used to trace the
transparent star-forming  \ion{H}{II} regions, where the presence
of dust is negligible and molecular gas has been wiped out
\citep{osterbrock89}. There has been a discussion about the
problems of using the 8 $\mu$m emission dominated by PAHs as a
tracer of star formation (SF), and no final conclusion has been
reached \citep{calzetti07,wu05}.  Since PAHs have been found to be
associated with very young stars, it is likely that its emission
is a reliable SF tracer. The relation between SF and emission at
24 $\mu$m due to a hot dust phase has been investigated several
times \citep[see][]{helou00}, and is found to originate in very
small dust grains (VSG) in regions surrounding the ionized gas.
Finally, CO traces the molecular clouds inside which the process
of SF occurs (molecular cores) or will occur \citep{kennicutt07}.

\subsection{On the value of the slope $\beta$}

The luminosity of an emitting cloud of gas (or even stars) is to a
first approximation given by
$L_{\nu}\propto\intop_{V}\epsilon_{\nu} dV$, where the integral is
computed over the entire volume of the cloud $V$ and
$\epsilon_{\nu}$ is the local emissivity at frequency $\nu$. If
all parameters that control $\epsilon_{\nu}$ are constant within
the volume, the expression reduces simply to $L_{\nu}\propto V$.
If the shape of
the cloud is globally spherical, 
then $L_{\nu}\propto V$ $\propto R^{3}$. In this case, we expect a
slope of $\beta=1$, which is observed in the H${\alpha}$ image of
\galb\ and in the NUV image of the same galaxy.

\subsubsection{The case of $\beta>1$}\label{beta}

\begin{figure*}[!]
  \resizebox{\hsize}{!}{\includegraphics{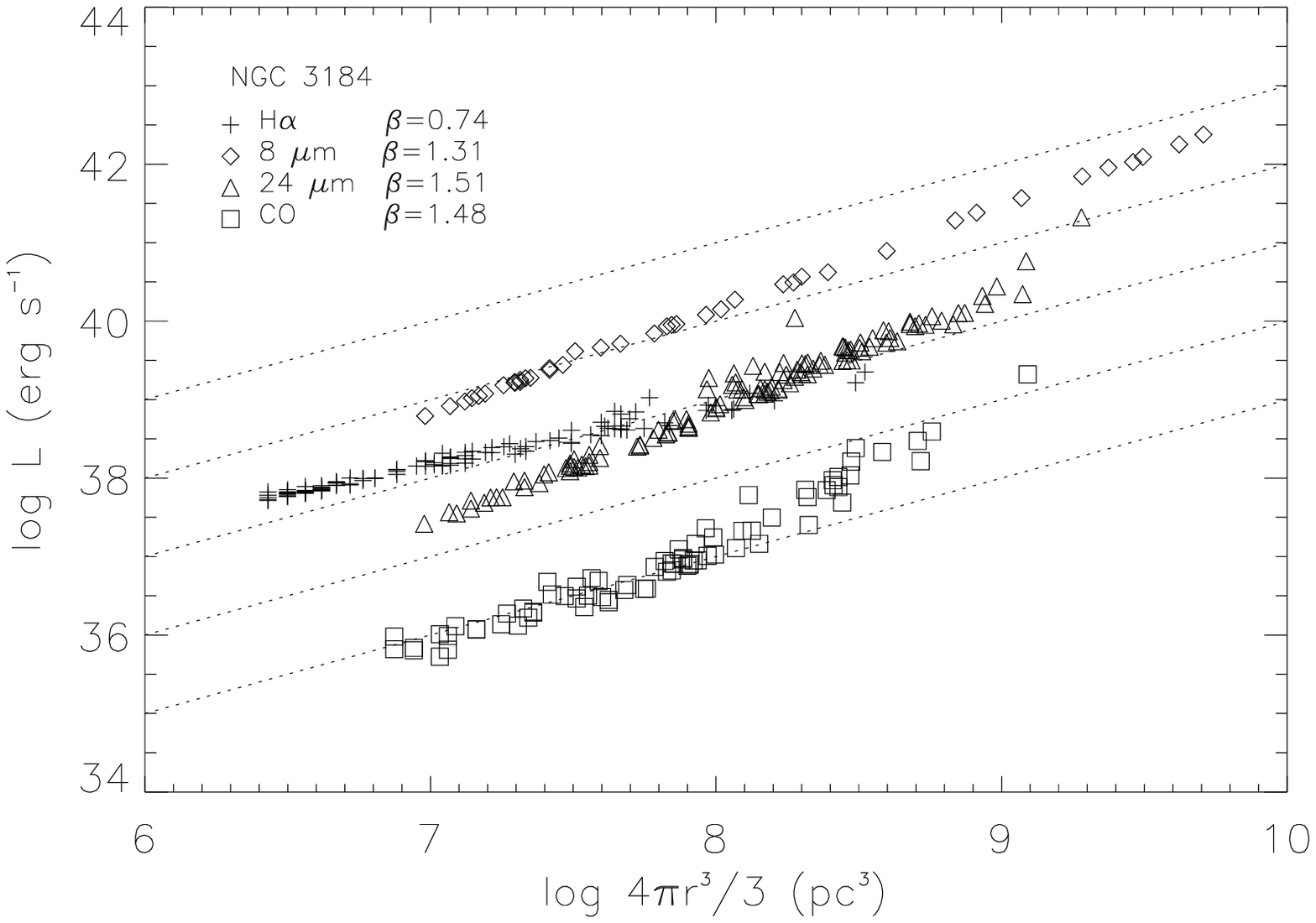}\includegraphics{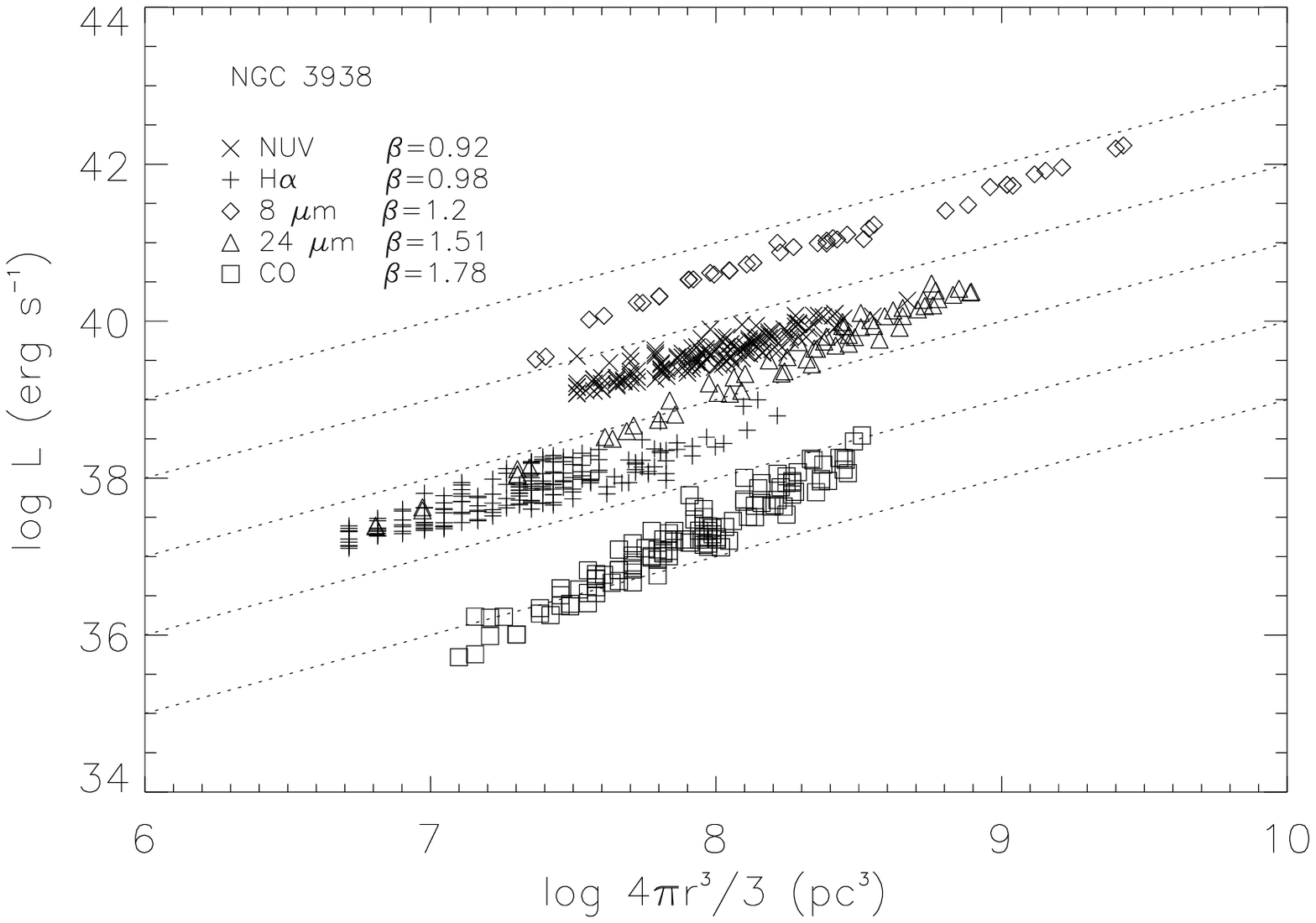}}
  \caption{Relation between the logarithm of the luminosity and the logarithm of the
  volume of the emission region for different star-formation indicators in \gala and \galb, $L_{\nu} \propto V^{\beta}$, where $V=4\pi r^3/3$ for $r$ the
  mean radius of the emission region. The slopes $\beta$ obtained for the different tracers
  are given in the upper left corner of the figures.}
  \label{fig3}
\end{figure*}

To derive a value of $\beta$ that differs from one, one of the
previous hypotheses must be changed:

\begin{itemize}
\item The shape must not be spherical: the clouds could be instead
elongated along the line of sight, and this elongation should
increase with an increasing size of the cloud. This hypothesis,
however, appears very unrealistic.

\item The local emissivity $\epsilon_{\nu}$ could vary with the
size of the cloud. This is possible if any parameter to which the
emissivity is related varies as well, for example the density of
the emitting gas. Most of the emissivity coefficients are related
to the square of the gas density. If this is the case, we could
obtain a slope larger than one if the mean density of the gas
increases with the volume of the cloud. This could be due to a
global increase in the density, or to the presence of high density
clumps, with high local emissivity coefficients. For these clumpy
regions, the resulting luminosity can be decomposed into two
contributions: $L_{\nu}\propto\epsilon_{\nu}^{0}
(V_{T}-V_{c})+\epsilon_{\nu}^{c} V_{c}$, where $V_{T}$ and $V_{c}$
are the total volume of the cloud and the volume of the high
density clumps, respectively. The emissivity coefficient of the
background low density gas is $\epsilon_{\nu}^{0}$ and the clump
emissivity coefficient is $\epsilon_{\nu}^{c}$. If the volume of
the clumps is related to the total volume of the cloud according
to $V_{c}=k V_{T}$, where $k$ is similar to a filling factor, we
can describe the luminosity by $L\propto[\epsilon_{\nu}^{0}
(1-k)+\epsilon_{\nu}^{c} k] V_{T}$. To observe any effect of the
clumps, the contribution of both phases of the gas should have the
same order of magnitude: $\epsilon_{\nu}^{0}
(1-k)\sim\epsilon_{\nu}^{c} k$. Since the clumps have higher
emissivity coefficients ($\epsilon_{\nu}^{c} >
\epsilon_{\nu}^{0}$) and a smaller relative volume ($k$ small),
one of these two parameters must increase with the size of the
cloud to obtain a luminosity higher than that expected from pure
volumetric effects ($\beta>1$). The first option is that the
emissivity of the dense clumps should increase: we can always
consider denser clumps as the total size of the cloud increases.
The second option is that the relative volume of the clumps
increases, i.e. there are more or larger clumps in larger clouds.
This would also explain the case of $\beta<1$ for H$\alpha$ in
\gala (see below).

\item It is known that due to density inhomogeneities, molecular
gas tends to form clumps. However, clumps found at mid-infrared
(MIR) emission regions (which could be explained by newly formed
high to intermediate mass stars embedded into the dust that heat
and re-emit with a temperature dependence $T^4$) are probably due
to a global increase in the density, or to the presence of high
density clumps, with high local emissivity coefficients. The
clumpy nature of the dusty regions is clearly revealed in both
galaxies. As shown in Fig.~\ref{fig1}, MIR clumps closely follow
the spiral arms and reach far larger volumes than the other
tracers (see Figs.~\ref{fig2} and \ref{fig3}).  The ionizing
regions detected by their H$\alpha$ and NUV emission appear to
span a smaller range in volume than  dusty and molecular gas
regions, which reach far larger volumes.

\end{itemize}

\subsubsection{The case of $\beta<1$}

The same study of the hypothesis corresponding to $\beta=1$ must
be completed to understand the possible situations leading to
$\beta<1$ (which was obtained for the H$\alpha$ image of \gala):

\begin{itemize}
\item The shape of the cloud is not spherical: it must smoothly
change from being spherical to oblate ellipsoidal, with a
decreasing size in the direction of the line of sight. In the
present case of face-on galaxies, this direction is perpendicular
to the plane of the galaxy. The value of $\beta$ would change from
1 to 2/3, as the luminosity would become proportional to the area.
This is not what is actually observed, since the slope is rather
constant for a large range of cloud sizes.

\item Another way of changing the shape of the emitting cloud is
to take into account the optical depth along the line of sight,
which is related to the cloud's size in this direction. Increasing
the size of the cloud would lead to high values of the optical
depth, such that the most distant part of the cloud could not be
seen. In this case, it would be the shape of the visible cloud
that would change, from a naturally spherical shape to a more
disk-shaped when observed face on. We would observe a type of
saturation effect as the optical depth started to become
important, which would again change the slope from 1 to 2/3,
something that is not observed.

\item The volume of the emitting cloud can also be reduced by
removing its inner part. This may be the case if there were some
places inside the cloud from which the gas does not emit. This
could be explained if some clumps of gas are in a phase in which
they do not emit. The presence of neutral or molecular high
density insertions (high density clumps inserted on less dense
surroundings), which not only do not emit in H$\alpha$, but also
conceal the gas behind them, can lead to a decrease of the
emitting volume as the size of the cloud increases. Such
insertions would also explain the $\beta>1$ slope in the case of
dust or CO maps as is observed (see Sect. \ref{beta}).

\item We did not explore the possibility that emissivity decreases
with the size of the cloud, since we cannot identify any physical
process that could be responsible for this behavior.
\end{itemize}

\section{Conclusions}

We have studied star-forming regions in the face-on spiral
galaxies \gala\ and \galb, which have emission detectable by a
variety of different tracers: NUV, H$\alpha$, CO, and 8 and 24
$\mu$m. By analyzing the multiwavelength luminosities $L_{\nu}$ of
star-forming emission regions and their corresponding volumes $V$,
we have found a relation of the form $L_{\nu} \propto V^{\beta}$.
If we suppose that the luminosity is directly proportional to the
volume, we should expect a value of $\beta=$1, which is indeed
observed for the images in H$\alpha$ and NUV in \galb.

However, from the CO and 8 and 24 $\mu$m images, a value of
$\beta\,>\,\rm{1}$ is derived for both \gala ($\beta_{CO}
\sim$1.8) and \galb ($\beta_{CO} \sim$1.5). We conclude that this
could be due to a change in the local emissivity $\epsilon_{\nu}$
with the size of the cloud, either by an augmentation of the
overall density or because of the presence of high density clumps,
with high local emissivity coefficients. For both galaxies we
measure similar values of $\beta$ for the dust tracers, i.e.
$\beta \sim$1.2--1.3 for 8 $\mu$m and 1.51 for 24 $\mu$m. From
previous studies, the clumpy structure found for molecular gas and
H$\alpha$ was expected. The mid-infrared emission of dust (PAH and
VSG) heated by newly formed high to intermediate mass stars is
revealed as clumps as well throughout the galaxies.

For the H$\alpha$ image of \gala, a slope of $\beta<$1 is
obtained, which could be interpreted by the presence of neutral or
molecular high density clumps, which not only do not emit in
H$\alpha$ but also conceal the gas behind them. This  can lead to
a decrease in the emitting volume as the size of the cloud
increases.

For \gala\ and \galb, the volumes found for ionizing regions
(H$\alpha$  and NUV emission) appear to span a smaller range than
that found for dusty (8 and 24 $\mu$m) and molecular gas regions.

\bibliographystyle{aa} 
\bibliography{caldu} 

\begin{acknowledgements}
We thank the referee for his/her valuable comments. ACP
acknowledges a scholarship from Sistema Nacional de Investigadores
(CONACYT, Mexico) and ICG acknowledges support from CONACYT
(Mexico) research grant 48484-F. This research has made use of the
NASA/IPAC Extragalactic Database (NED) which is operated by the
Jet Propulsion Laboratory, California Institute of Technology,
under contract with the National Aeronautics and Space
Administration.
\end{acknowledgements}

\end{document}